# Selective excitation and imaging of ultraslow phonon polaritons in thin hexagonal boron nitride crystals.

Antonio Ambrosio,[1,2,3,‡,*] Michele Tamagnone,[4,‡] Kundan Chaudhary,[4] Luis A. Jauregui,[2] Philip Kim,[2] William L. Wilson,[1] Federico Capasso.[4,*]

[1]Center for Nanoscale Systems, Harvard University, Cambridge, Massachusetts 02138, USA

[2]Department of Physics, Harvard University, Cambridge, Massachusetts 02138, USA

[3]CNR-SPIN U.O.S. Napoli, Complesso Universitario di Monte Sant'Angelo, Via Cintia, 80126 – Napoli, Italy

[4]Harvard John A. Paulson School of Engineering and Applied Sciences, Harvard University, Cambridge, Massachusetts 02138, USA

[‡] These Authors contributed equally to this work

*Corresponding Author, email: ambrosio@seas.harvard.edu, capasso@seas.harvard.edu

Polaritons in 2D and van der Waals (vdW) materials have been investigated in several recent works as an innovative platform for light-matter interaction, rich of new physical phenomena.[1, 2, 3, 4, 5, 6, 7, 8, 9, 10, 11, 12, 13, 14, 15]

Hexagonal Boron Nitride (h-BN), in particular, is an out of plane anisotropic material (while it is in-plane isotropic) with two very strong phonon polaritons bands where the permittivity becomes negative. In the first restrahlen band (RS1, 780-830 cm$^{-1}$) the relative out of plane permittivity $\varepsilon_{\parallel}$ is negative, while in the second restrahlen band (RS2, 1370-1610 cm$^{-1}$) the relative in-plane permittivity $\varepsilon_{\perp}$ is negative.[7] Due to these optical properties, thin h-BN flakes support guided modes which have been observed experimentally both via far field and near field methods. [1, 3, 6, 7, 8, 9, 10, 14]

In the literature, the first mode of the RS2 band has been studied and imaged extensively. [6, 8, 9] Due to the lack of suitable laser sources instead, the first mode in the RS1 band is more difficult to be accessed and imaged with near-field techniques. The observations of the RS1 modes have been in fact limited to spectral analysis,[3,8] while proper imaging has not been achieved yet.

In this work, we show how selectively excite the more confined modes in the RS1 and RS2 bands. The supported guided modes have phase and group velocities respectively tens and hundreds of times slower than the speed of light. We also show the possibility of full hyperspectral nano-imaging of modes in RS1 band by means of photo-induced force microscopy (PiFM). Moreover, a direct comparison of (PiFM) and scattering-type near-field microscopy (s-SNOM) is obtained by imaging the modes of the RS2 band with both techniques implemented on the same platform.

**The possibility of addressing ultraslow (ultraconfined) polaritonic modes of h-BN crystal flakes together with the possibility of optical nano-imaging in both the restrahlen bands have many innovative aspects that can lead to unprecedented schemes for strong light-matter interaction, slow and confined light.**

Our sample consists of a h-BN flake exfoliated from a bulk h-BN crystal using adhesive tape and transferred on a gold coated substrate (silicon + 500 nm wet thermal oxide + 100 nm e-beam evaporated gold with 5 nm chromium for adhesion, Figure 1a). The crystal, shown in Figure 1b, has a thickness $h$ of 121 nm measured via Atomic Force Microscopy (AFM) (Figure 1c) and sharp edges that facilitate the SPM imaging as explained later.

Both RS1 and RS2 bands are in wavelengths ranges where the gold substrate shows very high and negative permittivity values (in the order of thousands). Therefore, the surface of gold can be accurately approximated as a perfect electric conductor (PEC), i.e. as a perfect mirror. As discussed below, this configuration has two main consequences: 1) It simplifies the modelling and allows developing a simple description of the propagating guided polaritons in the h-BN crystal; 2) For symmetry reasons only odd polaritonic modes are allowed.

In fact, according to the image theorem, our layered system is equivalent to one where the mirror is removed and all the structures in the half-space above it (the h-BN sample, sources and fields) are replicated and flipped in the half-space below (Figure 1 d). In addition, charges and electric fields in the bottom half space are inverted upon reflection. The resulting structure, in our case, is then equivalent to a single h-BN layer with twice the thickness ($2h$), suspended in air (Figure 1b). Maxwell's equation can then be solved in this equivalent structure with the aim of finding the dispersion of the guided modes of the h-BN sample. For a *2h* thick h-BN layer suspended in air, a simple and exact dispersion relation can be found analytically, as opposed to the case of a flake on gold, which is more complex (Supplementary Information). In particular, due to the symmetry with respect to the central plane (the mirror plane), all guided modes in the h-BN flake must be either even or odd. In fact, if the structure and the fields are flipped upside down, the electric field must be either left invariant (even mode) or it must be multiplied by -1 (odd mode).

The dispersion (derived in the Supplementary Information) of all the modes for a *2h* thick flake suspended in air is given by the following transcendental equations (see also[3, 6, 7]):

$$\frac{\gamma_{z1}\varepsilon_d}{\gamma_{z2}\varepsilon_\perp}\coth(\gamma_{z1}h) + 1 = 0 \qquad \text{for even modes} \qquad (1)$$

$$\frac{\gamma_{z1}\varepsilon_d}{\gamma_{z2}\varepsilon_\perp}\tanh(\gamma_{z1}h) + 1 = 0 \qquad \text{for odd modes} \qquad (2)$$

with:

$$\gamma_{z1} = k_0\sqrt{\frac{\varepsilon_\perp}{\varepsilon_\parallel}n_{eff}^2 - \varepsilon_\perp} \ , \quad \gamma_{z2} = k_0\sqrt{n_{eff}^2 - \varepsilon_d} \ , \quad k_0 = \frac{\omega}{c} \qquad (3)$$

where $\varepsilon_d$ is the relative dielectric permittivity of the superstrate (here it is air, $\varepsilon_d = 1$), $k_0$ is the free space wavenumber, $\omega$ is the angular frequency, $c$ is the speed of light, $n_{eff}$ is the effective index of each guided mode and $\gamma$ are the propagation coefficients. Importantly, these equations are exact, opposite to the case of dielectric substrates usually reported in literature that is only valid as an approximation holding for large wavenumbers.[6] By inspection of the equations, when $\varepsilon_\parallel$ and $\varepsilon_\perp$ have opposite signs, solutions can be found for arbitrarily large $n_{eff}$ values, which is typical of the hyperbolic modes in h-BN. Figure 1e represents graphically equations 1 and 2, where each point of the figure is colored in purple when the even modes equation is satisfied and in green when the odd modes equation is satisfied. The color brightness also represents the attenuation of the modes with darker curves corresponding to modes with larger propagation attenuation (highly confined modes).

The modes of Figure 1e are those in general possible for a h-BN flake in air. However, as anticipated previously, only odd modes (in green) will be physical modes of the original flake on gold. This is because the PEC symmetry (also known as E symmetry boundary condition) requires inversion of all electrical charges in the image half-space, since positive charges induce negative image charges and vice versa. This is represented in Figure 1f where the field for the first odd modes in both RS1 and RS2 are represented. Notice that odd-symmetry implies that the electric field must be orthogonal to the mirror and the magnetic field must be parallel to it. Moreover, the first mode of the RS2 band (usually observed and reported in literature) is forbidden in our configuration since it is an even mode. This means that in the RS2 band, at each energy, the first odd mode (number 2 in the picture) is expected to be detected. From Figure 1e it is also clear that such mode is more confined with respect to the first mode since the corresponding effective refractive index ($n_{eff}$) value is much larger. By simply placing the h-BN crystal on a gold substrate we selectively access more confined modes propagating in the crystal.

Figure 2 shows the results of the experimental detection of the first mode of the RS1 band. For this optical nano-imaging we used Photo-induced Force Microscopy (PiFM).[16,17,18,19] This setup is based on a commercially available platform (Molecular Vista) and a pulsed quantum cascade laser (QCL) (Block Engineering). This technique belongs to the recent approaches to show the potential of mechanical detection for optical nano-imaging.[20,21]
In our setup, the laser is P-polarized, the pulse width is 20ns and the light energy (wavelength) can be tuned between 795 and 1900 cm$^{-1}$. The repetition rate is instead adjusted as described below.

Different from aperture-type near-field microscopy (a-SNOM),[22,23,24,25,26] the laser is focused to the tip/sample region by means of a low NA parabolic mirror at 60° from the vertical to the sample surface (Figure 2a). The probe is a gold coated tip with about 300 kHz resonance frequency. In such setup, the feedback on the second eigenmode of the AFM cantilever, oscillating at $\omega_2$ (Figure 2 b) is used to stabilize the tip-sample distance during scanning and to extract the sample morphology. The first eigenmode of the cantilever (usually used for topography imaging in standard AFM operating in tapping mode) is excited by the interaction forces between the sample and the tip, induced by the external laser source. This complex interaction is resonantly detected in the cantilever mechanics when the laser intensity is modulated at a frequency containing the resonance frequency of the first eigenmode of the cantilever, typically the frequency sum or difference between the first and second eigenmode resonances. The origin of this light-driven interaction (force) is usually described in terms of an interaction force between a light induced dipole in the tip and its mirror image in the sample. It has then been shown to be representative of the dispersive part of the material polarizability (*material component*, Figure 2b)[34,35]. This component mostly depends on the optical properties of the topmost material and it is the only component contributing to the detected signal when no guided modes are supported by the sample. However, when phonon polaritons are present, a local spatial modulation of the electric field at the surface is also presen.t[28] This also affects the tip-sample interaction and may produce, by means of interference with the material component, characteristic features (fringes) in the optical image (*polariton components*). In our configuration, two polariton components are in principle possible: a *standing component* (Figure 2c) that produces fringes in the optical image with periodicity twice the phonon polariton wavelength. This component originates from the scattering of the AFM tip. It propagates to the edge of the flake where it is reflected back and finally scattered by the AFM tip; a *direct component* (Figure 2c) that has the actual periodicity of the polariton mode and originates from scattering at the h-BN flakes sharp edge. A more accurate description of each component appears below.

Figure 2 d-g show examples of PiFM imaging close to the edge on the flake at different incident light energy. The experimental periodicities here correspond to the direct component only and their values are those expected for the first mode in the RS1 band.

It is to be noted here that imaging by means of a scattering-type Near-field Optical Microscope cannot be performed in order to reproduce such images. This is due to the lack of continuous laser sources in the RS1 band energy range which is necessary for s-SNOM experiments implementing pseudoheterodyne schemes. In fact, so far, only imaging of single fringes (only the stronger one, closer to the edge) has been monitored and reported in s-SNOM using super-broad synchroton sources[8] or as a derivation of nano-Fourier Transform IR spectra.[3]

Moreover, the fact that the pulsed QCL source can be rapidly tuned in wavelength allows us to sweep the energy across the whole laser emission range at each point of the scanning. This allows collecting hyperspectral images that can be used to fully recover the polaritonic dispersion of our sample. Figure 2h shows the experimental dispersion derived by such imaging. From the data reported in Figure 2h it is also possible to calculate other interesting quantities. Figure 2i and 2j show the theoretical and experimental values of the effective phase and group refractive index of the first mode in the RS1 band. The experimental results are in excellent agreement with the theory. The phase and group indices have very high values and opposite signs (negative group velocity). The mode is then strongly confined, with group velocity approximately 500 times slower than light in free space. This value is completely consistent with previous analysis on such modes.[3]

For the imaging and the analysis of the modes in the second RS band (1370-1610 cm$^{-1}$) we used both s-SNOM and PiFM. For this study, a commercially available s-SNOM platform was used (Neaspec). PiFM measurements were made utilizing home-implemented detection electronics to extract the cantilever dynamics on the same optical platform. s-SNOM and PiFM images in the RS2 band were then obtained nearly simultaneously (one after the other) on the same microscope with the same AFM probe (a PtIr coated cantilever, 70kHz resonance frequency) and the same laser, a QCL laser (DayLight). The laser is operated in continuous emission (CW) for s-SNOM and intensity modulated (50/50 duty cycle) for the PiFM imaging. This approach enables direct comparison between PiFM and s-SNOM imaging of the polaritonic modes.

Figure 3 a-d show four s-SNOM scanning (Amplitude maps from the pseudoheterodyne configuration) at different energies in the RS2 band. These images are used to calculate the group and phase indices in the RS2 band. According to the aforementioned theory, for our h-BN sample structure, in the RS2 band we expect the first mode to be forbidden since it is even. In fact, in both PiFM (Figure 3 f) and s-SNOM (figure 3g) images, the periodicity of the observed fringes corresponds to the second (first odd) mode of the RS2 band. From these images, it is clear that we observe both the double periodicity of the standing component and the single periodicity of the direct component, that actually becomes the only one visible far from the flake's edge. PiFM and s-SNOM (Amplitude map) show peaks in the same position for both the standing and the direct component (Figure 3h). This result is expected since both the PiFM signal and the s-SNOM signal (Amplitude) are described to be representative of the real part of the local polarizability (dispersive behavior).[27,28,29,30,31,32,33,34,35] In line with such description, the polaritonic mode represents only an extra modulation of the local field at the sample surface (or alternatively a local spatially modulated effective polarizability).

However, the s-SNOM setup combined with the CW QCL and the pseudoheterodyne detection allows retrieving also the local field phase rather than just the amplitude. Mapping both the amplitude and phase is crucial to get a more comprehensive picture of the polaritonic components (see below).

The existence of both direct and standing components, has been discussed for flat h-BN samples but never imaged so far. Our imaging sensitivity instead allow us to observe the interference fringes resulting from both these components. The absence of a standing component while imaging the mode in the RS1 can be explained considering that the electric field distribution of the polaritonic modes in RS1 band is more "internal" than in the RS2 band, with large field components deeper inside the material (Figure 1).[8] This condition reduces the coupling efficiency (in both injection and detection) for the modes launched by the external tip. In addition, the RS1 mode has a vertical electric field distribution all across the sample thickness (Figure 1). This also couples well with an incident wave at the edge of the crystal (direct mode). In contrast, the RS2 has a more complex field distribution, which reduces the edge coupling efficiency and makes the standing and direct components similar in terms of excitation and detection efficiency.

In more detail, what we call *material component* is the signal resulting from the local polarizability of the material. As such, it can be represented by a complex number $\tilde{A}_M$ whose strength depends by the local material response under the AFM tip and the phase depends by the relative position of the AFM tip with respect to the external illumination. The AFM tip is fixed with respect to the illuminating light, so the material component is expected to have the same complex value everywhere on a certain material, e.g. on the h-BN flake. In a complex plane representation, the complex values of the material component should distribute themselves around a single value.

The *direct component* is due to the phonon polariton wave launched by the flake edge. It propagates from the edge to the tip where it is scattered. This component can be represented by a complex number $\tilde{A}_D = A_D\, e^{-ik_{PP}x}$, where $k_{pp}$ is the phonon polariton wavenumber. The direct component spatially modulates the field in the medium which interferes with the material component under the tip to create intensity maxima and minima (fringes) with periodicity $\lambda_{pp}$ (phonon polariton wavelength). The amplitude of the direct component under the AFM tip is set by the internal coupling efficiencies and propagation losses while the phase depends by the distance $x$ between the edge and the AFM tip. As any other interference between a constant reference (in this case represented by the *material component*) and a propagating wave, the representation in the complex plane (*Re*, *Im*) corresponds to a spiral with the internal point set by the most delayed configuration (tip far from the flake edge).

The *standing component* is actually not a standing wave even though its periodicity (twice the direct component) is the same of a standing wave between the tip and the edge (here we use such name for continuity with the literature). The standing component is the phonon polariton launched by the AFM tip that comes back to the AFM tip after propagation and reflection at the flake edge, and it changes as $\widetilde{A_S} = A_S x^{-\frac{1}{2}} e^{-i2k_{PP}x}$ with propagation. In such scheme, if the AFM tip moves of $L$ towards the edge during scanning, the optical path in a round trip is *2L*. This originates the half periodicity observed for this component. Analogously to the direct component, the observed fringes originate from the interference of the *standing component* and the *material component* under the tip. The representation of the interference in the complex plane is again a spiral that converges to the center as that representing the direct component.

The fact that the s-SNOM setup based on pseudoheterodyne detection allows the simultaneous imaging of both the local field amplitude and phase makes it possible to retrieve each individual component. This is showed in Figure 4.

Figure 4a shows the amplitude ($A$) and phase ($\varphi$) of the s-SNOM second harmonic signal close to the flake edge. These values can be combined together and represented in a complex plane ($Re, Im$) = ($A \cos\varphi, A \sin\varphi$) (Figure 4b). In such representation, the material component produces a distribution of close-together values as described above. The direct and the standing components instead contribute with values distributed along the arms of two spirals. From such distribution of values, each single component can be retrieved as described below.

The spatially averaged (along $y$) profiles for amplitude, phase, real and imaginary part of the polaritonic wave are obtained from the maps in Figure 4a. These plots are showed in Figure 4c and d. Note that at a certain distance from the flake edge, the single periodicity of the direct component is the only one visible. This is expected since the direct component is attenuated less in propagation than the standing component that results from a wave that has propagated twice as much for each tip position. The profiles reported in Figure 4 c and d can be used to calculate the complex Fourier transform showed in Figure 4 e. This spectrum shows two peaks corresponding to the single and double periodicities of the direct and the standing component. Peaks appear only in the positive wavenumbers region, directly corroborating the description of both the direct and standing component as propagating waves, also propagating in the same direction (away from the edge).

Once the peaks are found, two bandpass filters can be applied to restrict the spectrum region (dashed areas of the spectrum) around the peaks and to retrieve the profiles of the direct and standing components by means of the inverse Fourier transform (Figure 4f). If the two components are now represented in the complex plane, clean spirals are found as expected. Figure 4 g and h show the phasor evolution (spiral) of the electric field resulting from the direct and standing components respectively.

In conclusion, we proposed a simple configuration that allows to selectively excite and detect only more confined phonon polaritons modes in h-BN crystals by forbidding the less confined modes by symmetry of sources.

We experimentally retrieved the mode dispersion in the first RS band by means of hyperspectral imaging of the modes with PiFM. We experimentally demonstrated that the first mode of the second RS band is suppressed in our sample and measured the second (first odd) mode dispersion by both s-SNOM and PiFM implemented on the same platform. This also constitutes an unprecedented comparison between s-SNOM and PiFM with results in line with previous theoretical work about the nature of the signals provided by these two different techniques.

These results represent an important advancement in engineering of polaritonic modes in vdW materials as well as a fundamental milestone in high-resolution optical imaging.

## Acknowledgement

This work was performed in part at the Center for Nanoscale Systems (CNS), a member of the National Nanotechnology Coordinated Infrastructure (NNCI), which is supported by the National Science Foundation under NSF award no. 1541959. CNS is a part of Harvard University. This work was supported by the National Science Foundation EFRI 2-DARE program through Grant No. 1542807. M.T. acknowledges the support of the Swiss National Science Foundation (SNSF) grant no. 168545.

## Author contributions

M.T. designed and fabricated the multilayered sample and performed the theoretical simulations. A.A. developed and performed the experimental characterization with help from M.T., K.W and W.W. h-BN crystals were provided by P.K and L.J. Data were analyzed by M.T. and A.A. A.A., M.T. and F.C. wrote the paper with inputs from all authors. A.A., M.T. and F.C. led the project.

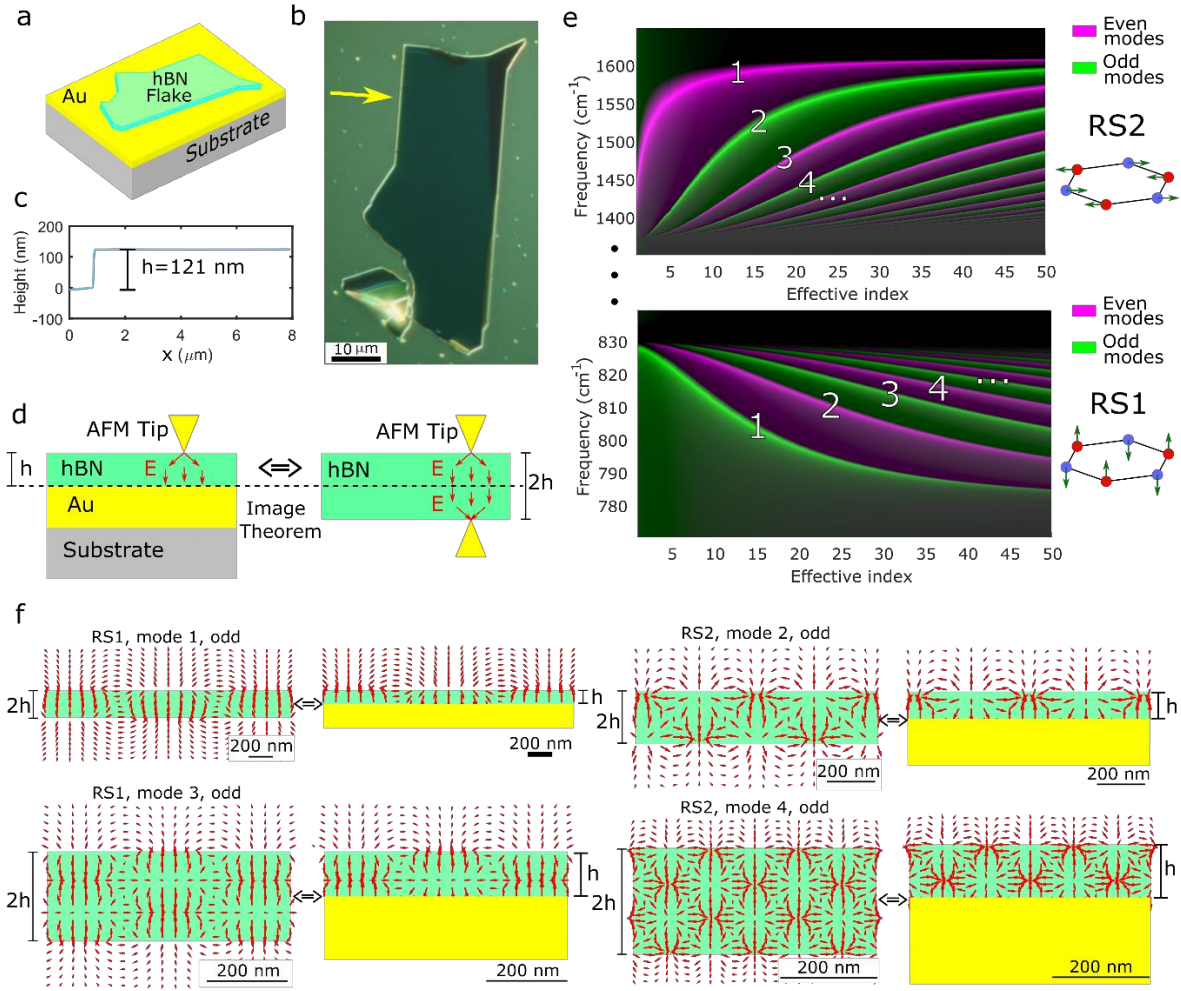

**Figure 1 | Selecting polaritonic modes by mirroring. a**, Schematic of our sample. A silicon substrate is coated with a 100 nm thick e-beam evaporated gold coating. A 121 nm thick h-BN flake is deposited over the gold film. **b**, Optical image of the h-BN crystallite using a phase contrast microscope. The yellow arrow points at the edge studied in this work. **c**, AFM profile of the flake at the edge. **d**, Equivalence between our multilayered sample and a single h-BN flake (twice as thick) in air, according to the *image theorem*. **e**, Polaritonic Mode dispersion calculated for a *2h* = 242 nm h-BN flake in air. In our case (E-symmetry) only odd guided modes are possible. **f**, Field

configuration for the first odd mode in each of the RS bands, and corresponding mode geometry in the h-BN on gold stack.

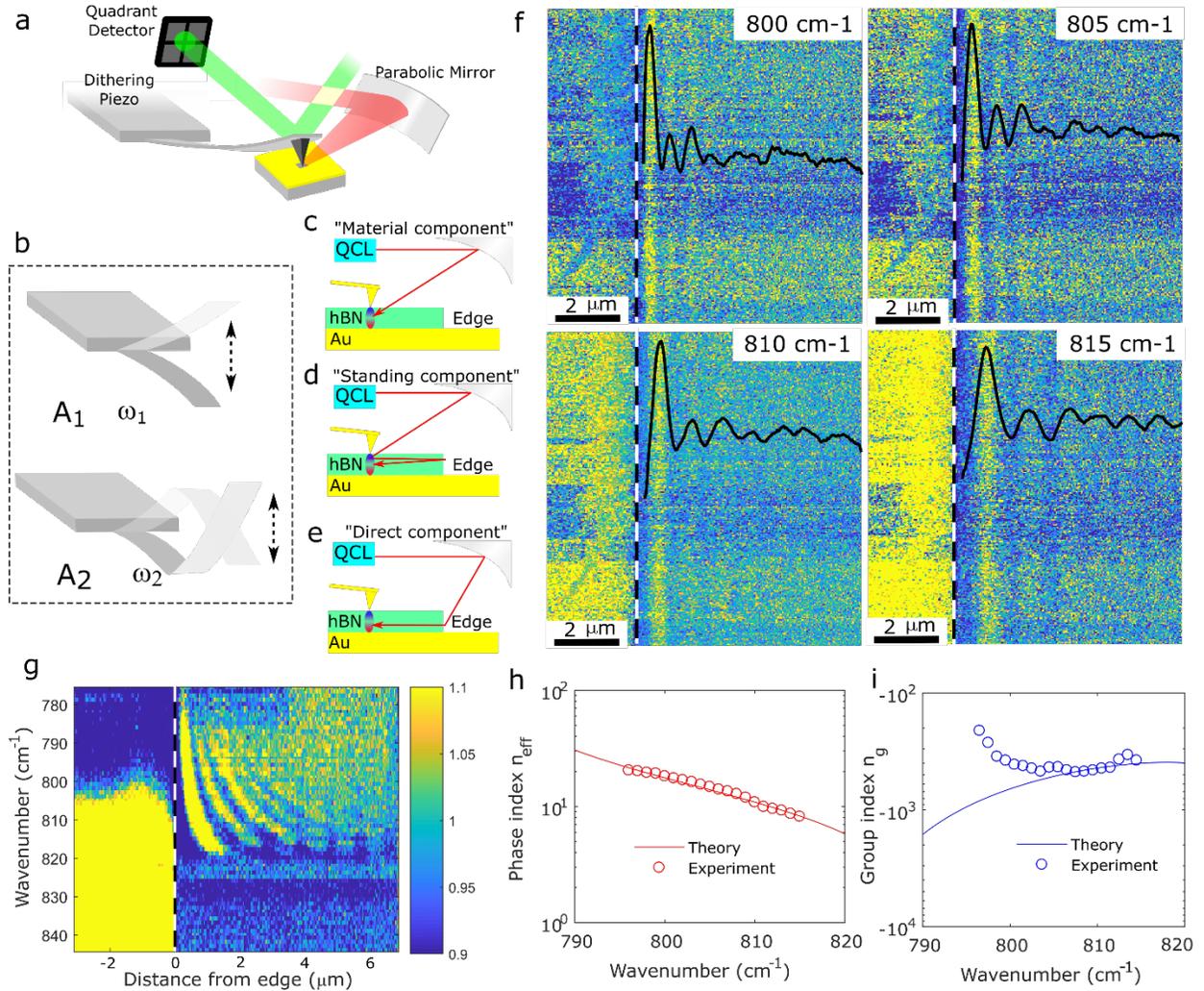

**Figure 2 | PIFM hyperspectral nano-imaging in the RS1 band. a**, Schematics of PiFM setup. **b**, First and second mechanical eigenmodes of an AFM cantilever. **c-e**, The three possible paths contribution to the measured signal. The material component (c) is due to the local dipolar response of the material. The standing component is due to a polaritonic guided mode originated by the AFM tip and that couples back with the tip after being reflected form the flake edge. The direct component originates from a polaritonic mode directly coupled into the flake by means of the scattering at the edge. **f**, PiFM imaging (and profile) of the fringes originated by the interference of the guided modes with the material component at the tip location for different energy (wavenumebers) of the illuminating light in the first RS band. **g**, Plot of the PiFM signal intensity

as a function of the distance from the edge and the illuminating light energy. This plot is obtained by averaging each of the hyperspectral set of data along the vertical axis (parallel to the flake edge) and it is used to retrieve the experimental mode dispersion in the h-BN crystallite and the data in (h) and (i). **h** and **i**, Analytical (Equation 2) and measured modal phase and group effective index as a function of illuminating light energy.

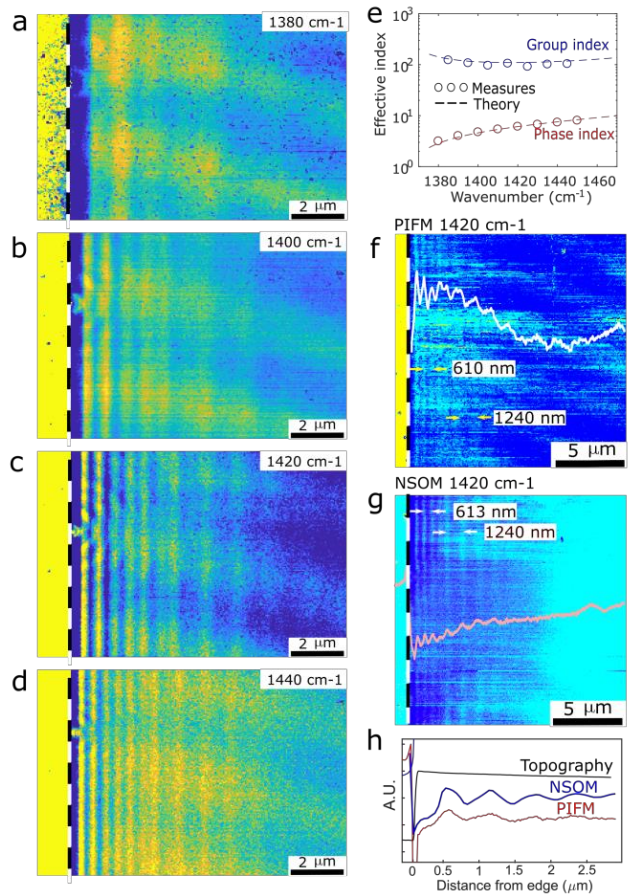

**Figure 3 | s-SNOM and PIFM images in the RS2 band. a-d**, s-NSOM images (Amplitude) of the h-BN polaritonic modes in the second RS band for different illuminating light energy. **e**. Group and phase index calculated from the s-SNOM data. **f-h**, Comparison between s-SNOM and PIFM imaging in the second RS band. Both the images show fringes with single (1240 nm) and double periodicity (610 nm), corresponding respectively to the direct and standing component of the first odd polaritonic guided mode. Both in PiFM and s-SNOM amplitude image, the fringes have maxima and minima in the same positions proving the similarity between the origins of the two signals.

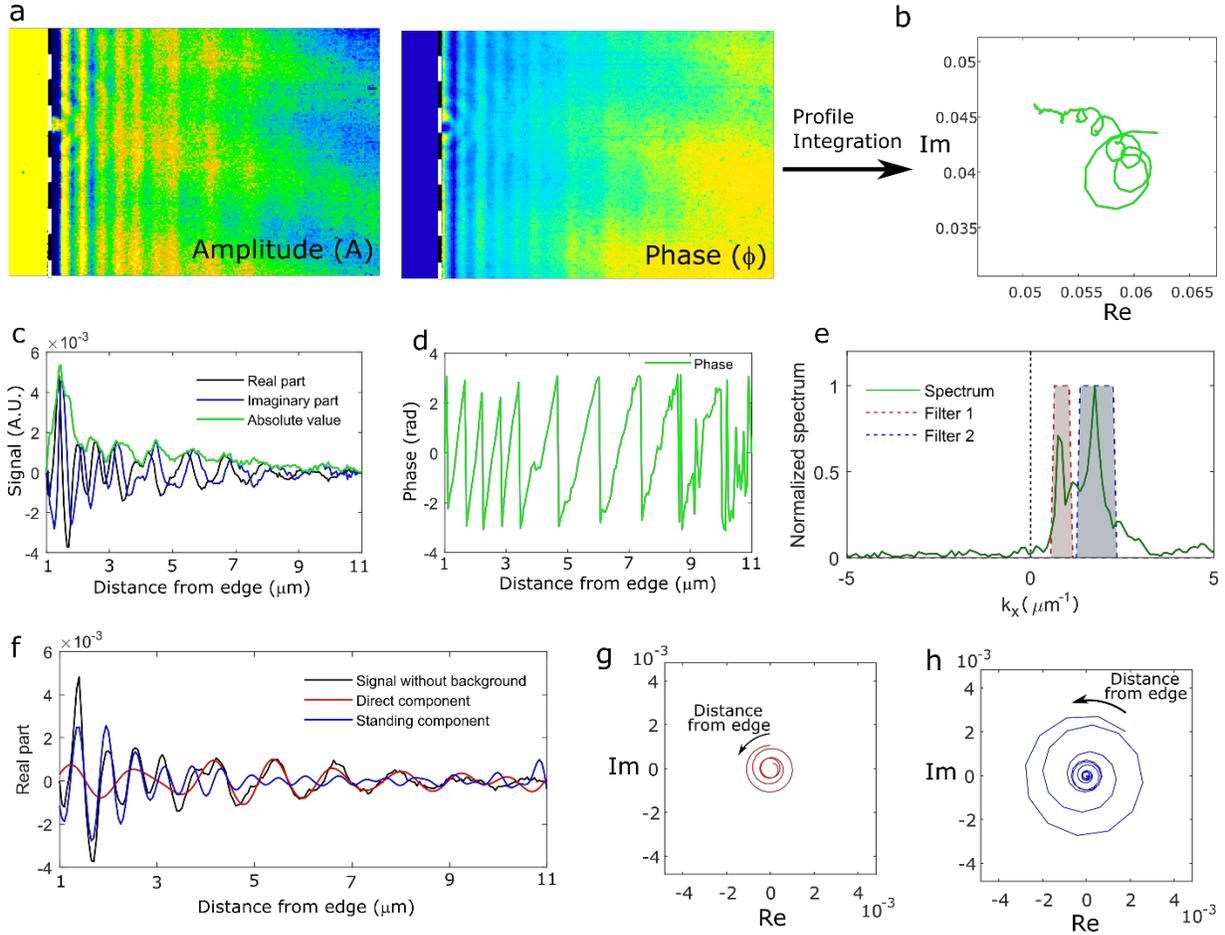

**Figure 4 | Extraction of the three polaritonic components with pseudoheterodyne s-SNOM. a** amplitude ($A$) and phase ($\varphi$) of the s-SNOM second harmonic signal close to the flake edge. These values can be combined together and represented in a complex plane ($Re, Im$) = ($A \cos\varphi, A \sin\varphi$) **b**. In such representation, the material component produces a distribution of close-together values; the direct and the standing components instead contribute with values distributed along the arms of two spirals. From such distribution of values, each single component can be retrieved as described below. **c** and **d** spatial average (along $y$) profiles for amplitude, phase, real and imaginary part of the polaritonic wave are obtained from the maps in **a**. **e** complex Fourier transform obtained form profiles reported in **c** and **d.** This spectrum shows two peaks corresponding to the single and double periodicities of the direct and the standing component.

Peaks appear only in the positive wavenumbers region, directly corroborating the description of both the direct and standing component as propagating waves, also propagating in the same direction (away from the edge). **f** inverse Fourier transform of **e** after applying two bandpass filters to restrict the spectrum region (dashed areas of the spectrum in **e**). **g** and **h** show the phasor evolution (spiral) of the electric field resulting from the direct and standing components respectively.

# Supplementary information

**h-BN modelling**

h-BN is modelled as an anisotropic material with the following relative permittivity tensor:

$$\bar{\bar{\varepsilon}}_r = \begin{pmatrix} \varepsilon_x & 0 & 0 \\ 0 & \varepsilon_y & 0 \\ 0 & 0 & \varepsilon_z \end{pmatrix}, \quad \varepsilon_x = \varepsilon_y$$

$$\varepsilon_x = \varepsilon_y = \varepsilon_\perp = \varepsilon_{\infty\perp}\left(1 - \frac{(\omega_{LO,\perp})^2 - (\omega_{TO,\perp})^2}{\omega^2 - j\omega\Gamma_\perp - (\omega_{TO,\perp})^2}\right)$$

$$\varepsilon_z = \varepsilon_\parallel = \varepsilon_{\infty\parallel}\left(1 - \frac{(\omega_{LO,\parallel})^2 - (\omega_{TO,\parallel})^2}{\omega^2 - j\omega\Gamma_\parallel - (\omega_{TO,\parallel})^2}\right)$$

$$\omega_{TO,\perp} = 1370\ cm^{-1} = 7.299\mu m = 170\ meV = 41.07\ THz \qquad (S1)$$

$$\omega_{LO,\perp} = 1610\ cm^{-1} = 6.211\mu m = 200\ meV = 48.27\ THz$$

$$\varepsilon_{\infty\perp} = 4.87 \qquad \Gamma_\perp = 5 cm^{-1}$$

$$\omega_{TO,\parallel} = 780\ cm^{-1} = 12.82\mu m = 96.7\ meV = 23.38\ THz$$

$$\omega_{LO,\parallel} = 830\ cm^{-1} = 12.048\mu m = 102.9\ meV = 24.88\ THz$$

$$\varepsilon_{\infty\parallel} = 2.95 \qquad \Gamma_\parallel = 4 cm^{-1}$$

These expressions can be cast into a Lorentz model with the following parameters:

$$\varepsilon_x = \varepsilon_y = \varepsilon_\perp = \varepsilon_{\infty\perp} + \frac{\varepsilon_{Lorentz\perp}(\omega_{0\perp})^2}{(\omega_{0\perp})^2 + j\omega 2\delta_\perp - \omega^2}$$

$$\varepsilon_z = \varepsilon_\parallel = \varepsilon_{\infty\parallel} + \frac{\varepsilon_{Lorentz\parallel}(\omega_{0\parallel})^2}{(\omega_{0\parallel})^2 + j\omega 2\delta_\parallel - \omega^2}$$

$$\omega_{0,\perp} = 1370\ cm^{-1} = 7.299\mu m = 170\ meV = 41.07\ THz = 258\ Trad/s$$

$$\varepsilon_{\infty\perp} = 4.87 \qquad \varepsilon_{Lorentz\perp} = 1.8557$$

$$\delta_\perp = 2.5 cm^{-1} = 0.075\ THz = 0.47\ Trad/s$$

(S2)

$$\omega_{0,\parallel} = 780\ cm^{-1} = 12.82\mu m = 96.7\ meV = 23.38\ THz = 146.9\ Trad/s$$

$$\varepsilon_{\infty\parallel} = 2.95 \qquad \varepsilon_{Lorentz\parallel} = 0.3903$$

$$\delta_\parallel = 2 cm^{-1} = 0.0600\ THz = 0.38\ Trad/s$$

**Guided modes in a suspended h-BN slab**

This section derives the field geometry and effective indexes of the modes in a h-BN slab with thickness $2h$ embedded in a dielectric with permittivity $\varepsilon_d$ (in our case air has $\varepsilon_d = 1$, but the symbol will be preserved for completeness). We assume without loss of generality that the modes are propagating in the in-plane x direction, with a guided wavenumber $k_x = n_{eff} k_0$ to be determined.

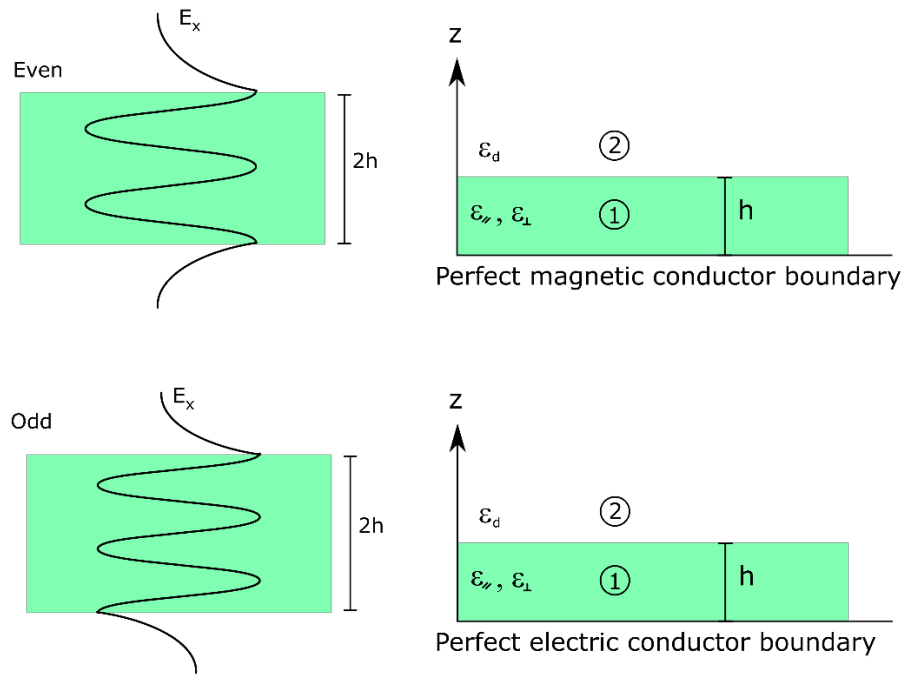

**Supplementary Figure 1 | Analytic derivation of modes dispersion.** Using the image theorem even and odd modes in an h-BN slab are equivalent to those found in a slab with half the thickness and terminated by perfect magnetic and perfect electric conductor boundary condition respectively.

Supplementary figure 1 shows the geometry of the problem. Using the image theorem, the even modes in a suspended h-BN slab are equivalent to those found in a slab with half thickness placed on a perfect magnetic conductor boundary condition. Similarly, odd modes are analogous to those in a half thickness slab on perfect electric conductor.

Assuming that $k_x$ is the guided wavenumber, then solving Maxwell's equations one can find two counterpropagating plane waves solutions in region 1 (h-BN) and one evanescent wave in region 2 (air). The magnetic and electric fields in region 1 are then given by:

$$\boldsymbol{H_1} = (0, H_{y1+}, 0)e^{\gamma_x x + \gamma_{z1} z} + (0, H_{y1-}, 0)e^{\gamma_x x - \gamma_{z1} z}$$

$$\boldsymbol{E_1} = \frac{iH_{y1+}}{\omega\epsilon_0}\left(\frac{\gamma_{z1}}{\varepsilon_\perp}, 0, -\frac{\gamma_x}{\varepsilon_\parallel}\right)e^{\gamma_x x + \gamma_{z1} z} + \frac{iH_{y1-}}{\omega\epsilon_0}\left(-\frac{\gamma_{z1}}{\varepsilon_\perp}, 0, -\frac{\gamma_x}{\varepsilon_\parallel}\right)e^{\gamma_x x - \gamma_{z1} z} \tag{S3}$$

with:

$$\gamma_{z1} = k_0\sqrt{\frac{\varepsilon_\perp}{\varepsilon_\parallel}n_{eff}^2 - \varepsilon_\perp} \; , \; \gamma_{z2} = k_0\sqrt{n_{eff}^2 - \varepsilon_d} \; , \; k_0 = \frac{\omega}{c} \tag{S4}$$

The evanescent wave in region 2 is described as:

$$\boldsymbol{H_2} = (0, H_{y2}, 0)e^{\gamma_x x + \gamma_{z2} z}$$

$$\boldsymbol{E_2} = \frac{iH_{y2}}{\omega\epsilon_0}\left(-\frac{\gamma_{z1}}{\varepsilon_d}, 0, -\frac{\gamma_x}{\varepsilon_d}\right)e^{\gamma_x x + \gamma_{z2} z} \tag{S5}$$

For odd modes $H_{y1+} = H_{y1-}$, while for even modes $H_{y1+} = -H_{y1-}$. Therefore, the fields in region 2 can be written as:

ODD CASE:

$$\mathbf{H_1} = 2(0, H_{y1} \cosh \gamma_{z1} z, 0) e^{\gamma_x x}$$

$$\mathbf{E_1} = \frac{2iH_{y1}}{\omega \epsilon_0} \left( \frac{\gamma_{z1}}{\varepsilon_\perp} \sinh \gamma_{z1} z, 0, -\frac{\gamma_x}{\varepsilon_\parallel} \cosh \gamma_{z1} z \right) e^{\gamma_x x} \quad (S6)$$

EVEN CASE:

$$\mathbf{H_1} = 2(0, H_{y1} \sinh \gamma_{z1} z, 0) e^{\gamma_x x}$$

$$\mathbf{E_1} = \frac{2iH_{y1}}{\omega \epsilon_0} \left( \frac{\gamma_{z1}}{\varepsilon_\perp} \cosh \gamma_{z1} z, 0, -\frac{\gamma_x}{\varepsilon_\parallel} \sin \gamma_{z1} z \right) e^{\gamma_x x} \quad (S7)$$

Applying the continuity of transversal magnetic and electric field at the interface between region 1 and 2 we find

ODD CASE:

$$2H_{y1} \cosh \gamma_{z1} d = H_{y2} e^{\gamma_{z2} d}$$

$$2H_{y1} \frac{\gamma_{z1}}{\varepsilon_\perp} \sinh \gamma_{z1} d = -H_{y2} \frac{\gamma_{z2}}{\varepsilon_d} e^{\gamma_{z2} d} \quad (S8)$$

EVEN CASE:

$$2H_{y1} \sinh \gamma_{z1} d = H_{y2} e^{\gamma_{z2} d} \quad (S9)$$

$$2H_{y1}\frac{\gamma_{z1}}{\varepsilon_\perp}\cosh\gamma_{z1}d = -H_{y2}\frac{\gamma_{z2}}{\varepsilon_d}e^{\gamma_{z2}d}$$

Dividing member by member:

ODD CASE:

$$\frac{\gamma_{z1}\varepsilon_d}{\gamma_2\varepsilon_\perp}\tanh\gamma_{z1}d + 1 = 0 \qquad (S10)$$

EVEN CASE:

$$\frac{\gamma_{z1}\varepsilon_d}{\gamma_2\varepsilon_\perp}\coth\gamma_{z1}d + 1 = 0 \qquad (S11)$$

which are the dispersion relations used in this work.

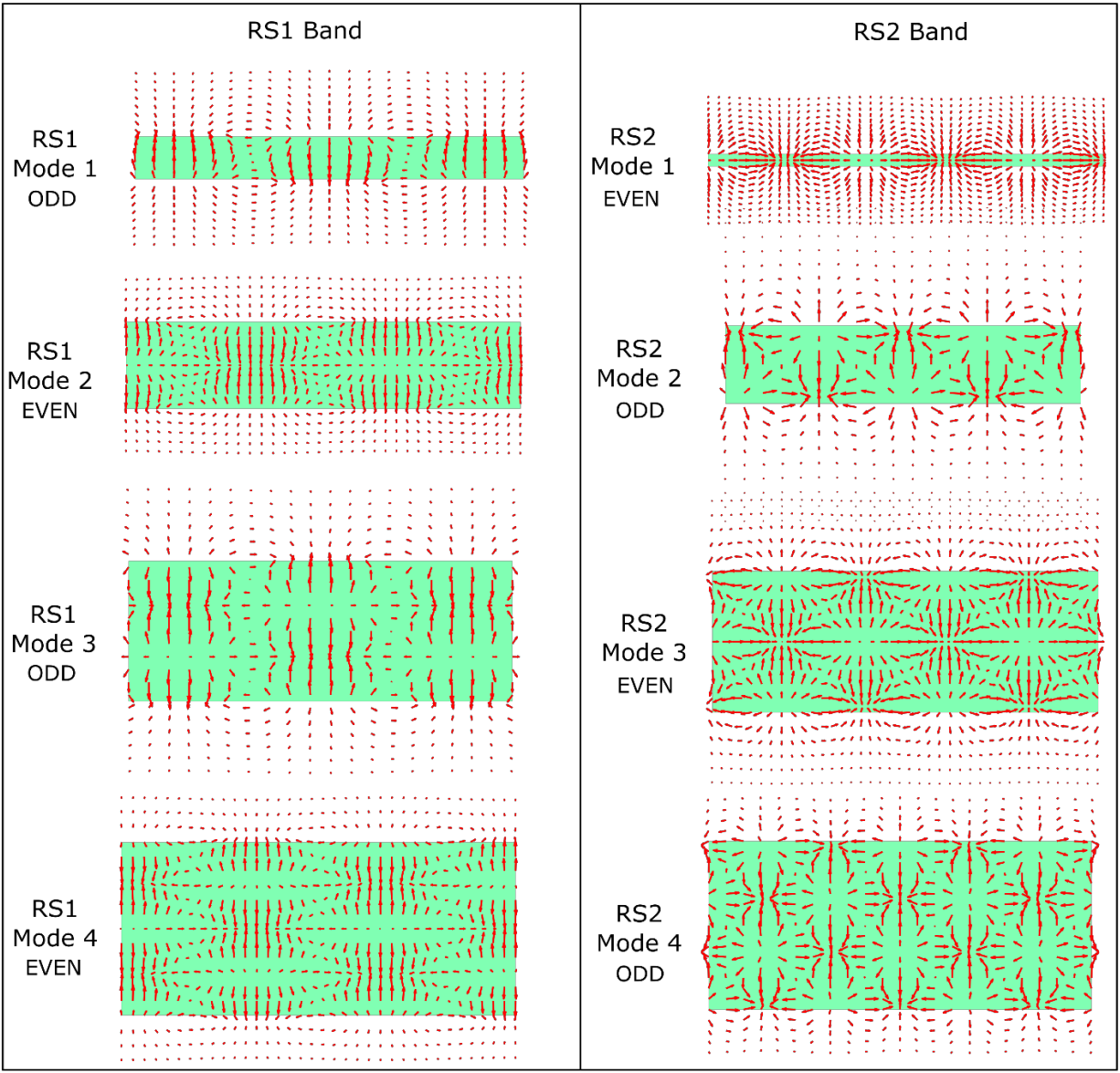

**Supplementary Figure 2 | Modes of suspended h-BN slab.** The first four calculated modes for a suspended h-BN flake are shown for each of the RS bands. All the modes propagate from left to right. RS1 fields are shown at the frequency of 818 cm$^{-1}$, while fields in the RS2 cases are shown at 1500 cm$^{-1}$. The thickness of the flake is $2h = 242\ nm$

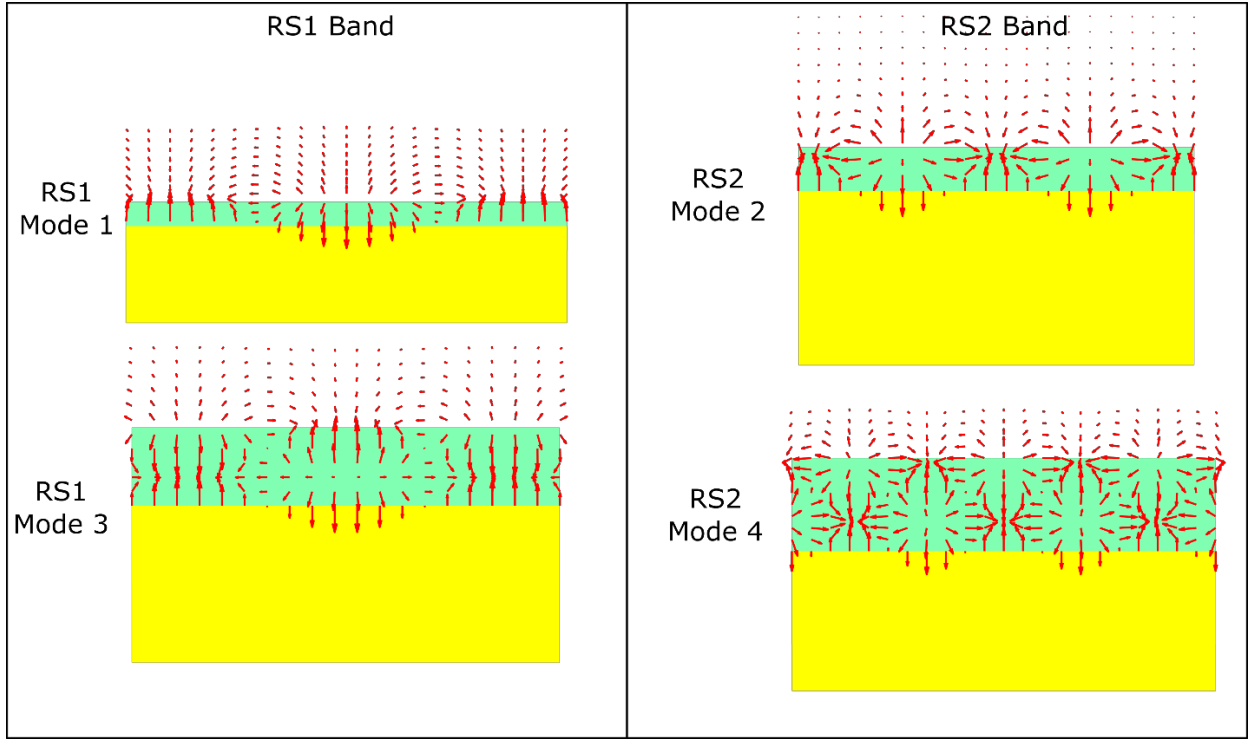

**Supplementary Figure 3 | Modes of h-BN on metal.** The first two allowed modes (odd modes) for the considered structure are shown for each of the RS bands. Plot parameters are the same as in Supplementary Figure 2, but the thickness of h-BN is $h = 121\ nm$

**Additional measurements**

This section introduces some additional measurements showing the presence of the interference fringes in the s-SNOM amplitude images done in the RS2 band with different harmonics and tip materials (Supplementary Figure 4). For metallic and silicon tips and for the second, third and fourth harmonics both the direct and standing interference patterns are observed. A defect is visible on the edge, and the circular waves launched by it are also visible.

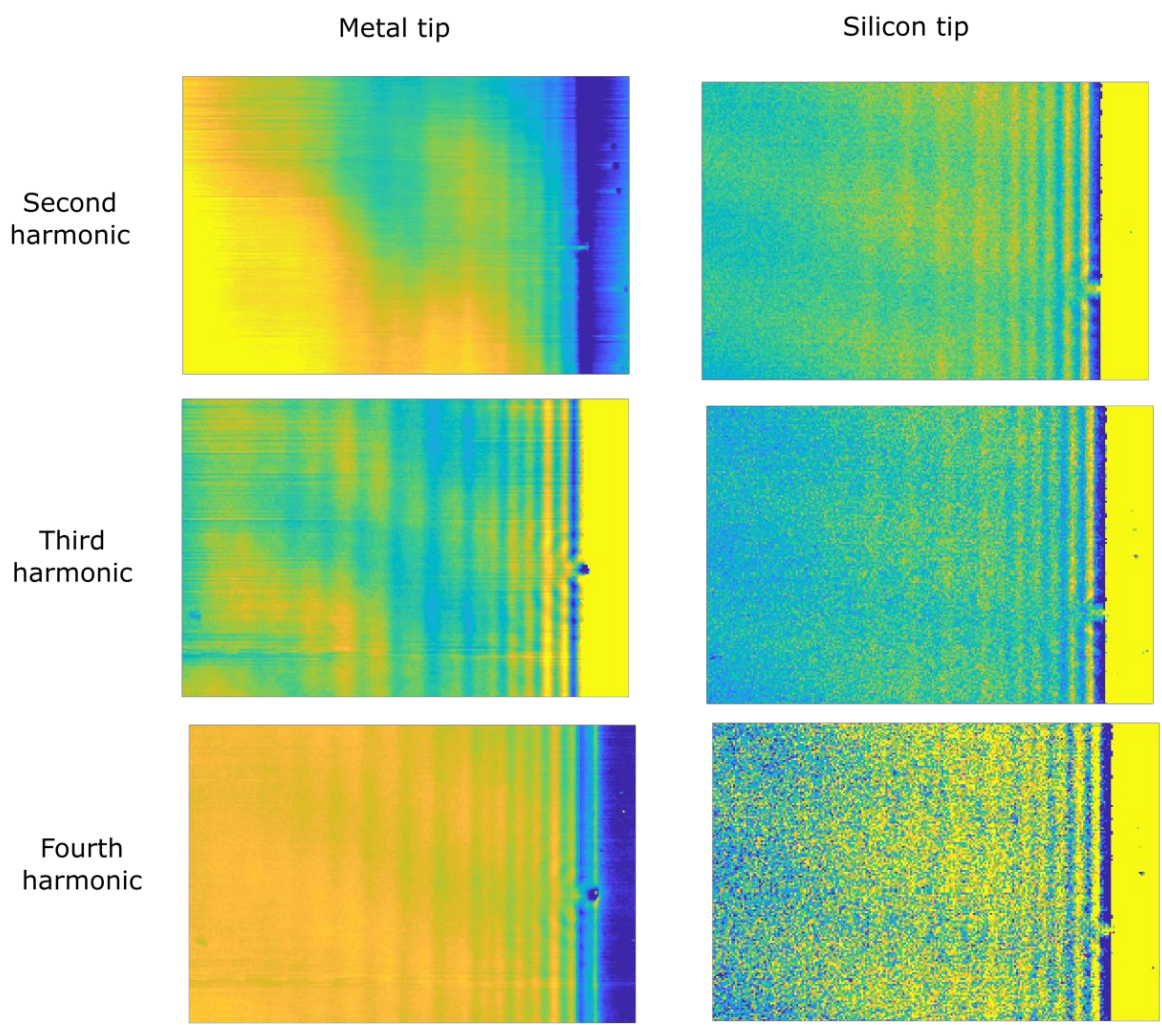

**Supplementary Figure 4 | s-SNOM images for different tip materials and harmonics.** All the plots represent the s-SNOM amplitude, measured at 1440 cm$^{-1}$.

**Effect of the beam inclination on the direct component**

The fringes of the direct component observed in the RS1 an RS2 bands are not exactly spaced as the guided wavelength of the mode. This is because the edge moves as the sample is scanned while the illuminating beam is tilted, and therefore there is a de-phasing of the launched mode. This dephasing can be corrected knowing the direction of the illuminating beam ($\theta = 60°$ with respect to the normal, $\phi = 45°$ with respect to the edge). Since the transversal wavevector ($k_x = \sin(\theta)\cos(\phi) k_0$) of the impinging wave then adds to the guided mode to form the fringes, then the correction consists in subtracting $\sin(\theta)\cos(\phi)$ from the measured effective index. However, such effect is small given the large effective indexes.